\documentclass[a4paper, 12pt, oneside]{article}
\usepackage[cp1251]{inputenc}
\usepackage[english, russian]{babel}
\usepackage{ graphics,amsfonts, amsmath,bm,amssymb, array}
\usepackage[dvips]{graphicx}
\usepackage[flushleft,small]{caption2}
\makeatletter

\newcommand{\const}{\mathop{\rm const\, }}
\renewcommand{\Re}{\mathop{\rm Re\,}}

\makeatother {

\begin{document}
\thispagestyle{empty} \large
\renewcommand{\abstractname}{\, }
\renewcommand{\refname}{\begin{center}REFERENCES\end{center}}

 \begin{center}
{ \textbf{Temperature jump in second Stokes' problem by nonlinear analysis}} \bigskip
\end{center}

\begin{center}
  \bf A. V. Latyshev\footnote{$avlatyshev@mail.ru$} and
  A. A. Yushkanov\footnote{$yushkanov@inbox.ru$}
\end{center}\medskip

\begin{center}
{\it Faculty of Physics and Mathematics,\\ Moscow State Regional
University, 105005,\\ Moscow, Radio str., 10A}
\end{center}\medskip \medskip

\begin{abstract}
The second Stokes problem about behavior of the viscous fluid (matter) filling
half-space is considered. A flat surface limiting half-space makes harmonious
oscil\-la\-tions in the eigen plane. The equations of mechanics of the continuous
environment are used. It is shown, that in considered process there is the
temperature difference between temperature of a surface and temperature environments
far from a surface. Usually similar difference is called as temperature jump. The
method of small parameter is applied. Temperature jump in the second approach is
found out.

\end{abstract}
\bigskip

\medskip

\bigskip \bigskip

For the first time the problem about behavior of gas over a wall oscillating in the
eigen plane, it is considered by G. G. Stokes \cite{Stokes}. It often called as the
second Stokes' problem.

Problem about behavior of the viscous fluid matter over a moving surface in the last
years the steadfast attention attracts \cite{Ai} -- \cite{Dudko}. It is
connected with development of modern technologies, in particular, nano-technologies.

Recently the problem about fluctuations of a flat surface  and
for a case non-Newton liquids \cite{Ai} is studied.

For a various sort of currents of environment the second Stokes' problem  was studied in \cite{Khan}, in \cite{Steinhell} questions of a friction and warmth allocation, in
\cite{Graebel} were investigated every possible appendices of this problem were considered.

In \cite{Cleland} the example of practical application of the oscillatory
The system similar to the considered problem, in area нанотехнологий is considered.

In experiments \cite{Karabacak} the stream of gas created by the mechanical
resonator at various frequencies of fluctuation was studied.
For a case of low frequencies the problem on the basis of the equation
of Navier---Stokes' is solved. In case of any speeds of fluctuations of a surface
were used numerical methods on the basis of the kinetic equation
with integral of collisions in the form of BGK (Bhatnagar, Gross and Krook).

In gases with use of the kinetic equations the second problem Стокса 
in works \cite{Karabacak} -- \cite{CMMP} was considered.

In work \cite{Sharipov} numerical methods of the decision of second Stokes' problem 
were applied, and in works \cite{Dudko0} and \cite{Dudko} for the problem decision were used
approximations methods. Second Stokes' problem in electromagnetic fields was
it is investigated in \cite{Choudhary}.

In our works \cite{MGG} and \cite{CMMP} the second Stokes' problem  analytically for
rarefied gas is solved. The model Boltzmann' kinetic equation 
with integral of collisions type BGK thus was used. 
In \cite{MGG} the problem with diffusion boundary conditions was solved. 
It is shown, that results of works \cite{Sharipov} and \cite{Dudko}
are rather close with the results received from the analytical decision \cite{MGG}.
Results from \cite{MGG}  on a case of boundary Cercignani' conditions 
in \cite{CMMP} have been generalized.

In the present work we find jump of temperature by which process
of generating of a sound wave by a surface oscillating harmoniously in the plane
is accompanied.

The method of small parameter is applied. Temperature jump is found out in the
second approximation.
\vspace{1cm}

Let the viscous fluid matter fills half-space $x> 0$ over the flat firm surface
laying in a plane $x=0$. The surface $ (y, z) $ makes harmonious oscillations along
an axis $y $ under the law $u_s(t)=U_0e^{-i\omega t} $. It is required to find jump
of temperature, formed between temperature of the wall and temperature of viscous
fluid matter far from a wall. 

We take system of the equations of the continuous
environment (see, for example \cite{Landau6})
$$
\rho\big(\dfrac{\partial V_i}{\partial t}+v_k\dfrac{\partial V_i}{\partial
x_k}\big)+\dfrac{\partial p}{\partial x_i}=
$$
$$
=\dfrac{\partial}{\partial x_k}\Big\{\eta \big(\dfrac{\partial V_k}{\partial x_i}+
\dfrac{\partial V_i}{\partial x_k}-\dfrac{2}{3}\delta_{ik}{\rm div}{\bf
V}\big)\Big\}+ \dfrac{\partial}{\partial x_i}\zeta ({\rm div}\bf V),
\eqno{(1)}
$$
$$
\rho T\big(\dfrac{\partial s}{\partial t}+{\bf V}\nabla s\big)=$$$$=
{\rm div}(\varkappa \nabla T)+
\dfrac{\eta}{2}\Big(\dfrac{\partial V_k}{\partial x_i}+\dfrac{\partial V_i}{\partial x_k}-
\dfrac{2}{3}\delta_{ik}{\rm div}{\bf V}\Big)^2+\zeta \big({\rm div}{\bf V}\big)^2,
\eqno{(2)}
$$
and
$$
\dfrac{\partial \rho}{\partial t}+{\rm div}(\rho { \bf V})=0.
\eqno{(3)}
$$\medskip

Here 
$s$ is the entropy of a mass unit of environment,
${\bf V}=(u,v)$ is the velocity of fluid matter, $u=u_y(t,x)$, $v=u_x(t,x)$ are components 
of velocity of fluid matter, $\eta$ is the 
dynamical viscosity of fluid, $\zeta$ is the second viscosity of fluid,  
$\varkappa$ is the heat conductivity coefficient, $T$ is the temperature of fluid matter, 
$c_p$ is the thermal capacity  at constant pressure, $p$ is the pressure in fluid matter.
For gases $p={\rho k_B T}/m$, $k_B$ is the Boltzmann constant, $m$ is the mass of molecule,
$\rho$ is the density of fluid matter.

On repeating indexes summation is meant.

Let us calculate the quantity
$$
\Big(\dfrac{\partial V_k}{\partial x_i}+\dfrac{\partial V_i}{\partial x_k}-
\dfrac{2}{3}\delta_{ik}{\rm div}{\bf V}\Big)^2=
$$

$$
=\Big(\dfrac{\partial V_k}{\partial x_i}+\dfrac{\partial V_i}{\partial
x_k}\Big)^2-\dfrac{4}{3}\delta_{ik}{\rm div}{\bf V}\Big(\dfrac{\partial V_k}{\partial x_i}+
\dfrac{\partial V_i}{\partial x_k}\Big)+\dfrac{4}{9}\delta_{ik}^2({\rm div}{\bf
V})^2=
$$
$$
=\Big(\dfrac{\partial V_1}{\partial x_2}+\dfrac{\partial V_2}{\partial
x_1}\Big)^2+\Big(\dfrac{\partial V_2}{\partial x_1}+\dfrac{\partial V_1}{\partial
x_2}\Big)^2+\Big(\dfrac{\partial V_1}{\partial x_1}+\dfrac{\partial V_1}{\partial
x_1}\Big)^2-
$$

$$
-\dfrac{4}{3}\Big(\dfrac{\partial V_1}{\partial x_1}\Big)\Big(\dfrac{\partial V_1}{\partial x_1}+
\dfrac{\partial V_1}{\partial x_1}\Big)+\dfrac{4}{9}(\delta_{11}+\delta_{22}+\delta_{33})
\Big(\dfrac{\partial V_1}{\partial x_1}\Big)^2=
$$

$$
=2\Big(\dfrac{\partial V_2}{\partial x_1}\Big)^2+4\Big(\dfrac{\partial V_1}{\partial
x_1}\Big)^2-\dfrac{8}{3}\Big(\dfrac{\partial V_1}{\partial
x_1}\Big)^2+\dfrac{4}{3}\Big(\dfrac{\partial V_1}{\partial
x_1}\Big)^2=
$$

$$
=2\Big(\dfrac{\partial u}{\partial x}\Big)^2+\dfrac{8}{3}\Big(\dfrac{\partial v}{\partial
x}\Big)^2.
$$

The boundary condition is formulated from the condition, that the wall makes, as
already it was specified, harmonious oscillations. Hence, the boun\-da\-ry condition
on the wall has the following form
$$
u(t,0)=U_0\cos \omega t=\Re(U_0 e^{-i\omega t}).
\eqno{(4)}
$$

In boundary conditions (4) possibility of sliding of gas along a surface is neglected. 
It means, that corresponding Knudsen' number it is not enough. 
For the considered problem it corresponds to the case of the low
frequencies. As shown in \cite{Sharipov} it corresponds to a condition
$$
\omega\ll \dfrac {\nu} {\lambda^2}.
$$

Here $ \nu =\eta/\rho $ is the coeffiecient of kinematic viscosity,
$ \lambda $ is the  length of free run of gas molecules.

In case of a liquid the specified parity (a sticking condition) is carried out automatically.

We consider as small parameter of the problem the quantity $ \varepsilon \;
(\varepsilon\ll 1) $, which is the relation of amplitude of velocity of the wall
$U_0$ to thermal velocity of environment $v_T =\sqrt {2k_BT/m} $, i.e. $
\varepsilon=U_0/v_T $. The velocity $v_T$ has an order of a sound velocity.

The problem (1) -- (4) we will solve by the method consecutive approxi\-ma\-tions.

As a first approximation on small parameter from considered system there is the 
equation (1) only.

As a first approximation on small parameter the problem has isothermal and isobaric character.
Concentration of gas as remains to a constant.
Thus from considered system there is an equation (1) only, which in this case
it will be transformed to the form
$$
\dfrac {\partial u} {\partial t} =
\nu\dfrac {\partial^2 u} {\partial x^2}.
\eqno {(5)}
$$

Other equations in the given approximation are carried out automatically.

For the decision of the equation (5) with a boundary condition (4) we search in the form
$$
u =\Re (U_0\exp (-i\omega t+k_1x)).
$$

We obtain that
$$
k_1=\sqrt{-\dfrac{i\omega}{\nu}}=\sqrt{\dfrac{\omega}{2\nu}}(i-1)=k(i-1),\quad
k=\sqrt{\dfrac{\omega}{2\nu}}.
$$

Therefore, the solution of equation (1) is constructed
$$
u=U_0\cos(-\omega t+kx)\exp(-kx)=U_0\Re(e^\xi),
\eqno{(6)}
$$
where $\xi$ is the phase of solution $\xi=-i\omega t-kx(1-i)$.

The derivative of the soltution (6) is equal
$$
\dfrac{\partial u}{\partial x}=\dfrac{U_0k}{2}\Big[(i-1)e^\xi-(i+1)e^{\xi^*}\Big].
$$

The derivative squared is equal
$$
\left(\dfrac{\partial u}{\partial x}\right)^2=\dfrac{U_0^2k^2}{4}
\Big[4e^{-2kx}-2ie^{2\xi}+2ie^{2\xi^*}\Big]=
$$
$$
=U_0^2k^2\Big[e^{-2kx}-i\dfrac{e^{2\xi}-e^{2\xi^*}}{2}\Big].
\eqno{(7)}
$$

Thus the first summand only  in the square brackets does not depend on time.

Further instead of a designation $v_2$ for $x$-component of speeds of environ\-ment 
in the second app\-ro\-xi\-ma\-ti\-on we use the designation $v $, as in the first
approximation this quantity is equal to zero.

Let us consider viscosity coefficient of environments to constants. 
In this case quantity  $u$ in the second approximation does not vary. 
We will notice, that this assumption not it essential to calculation of temperature jump.

Now the system (1) -- (3) in the second approximation can be copied in the form
$$
\rho\Big (\dfrac {\partial v}{\partial t} +v\dfrac{\partial v}{\partial x} \Big) =
 -\dfrac{\partial p}{\partial x} + \dfrac{\partial^2 {v}}{\partial  x^2} 
 \Big(\dfrac{4}{3} \eta +\zeta\Big),
\eqno {(8)}
$$
$$
\rho T\Big(\dfrac{\partial s}{\partial t} + v\dfrac{\partial s}{\partial x} \Big) =
\varkappa \triangle T + \dfrac{\partial^2 {v}}{\partial
 x^2} \Big(\dfrac{4}{3} \eta +\zeta\Big) + \eta\Big(\dfrac{\partial u}{\partial x} \Big)^2,
\eqno {(9)}
$$
and
$$
\dfrac{\partial \rho}{\partial t} =-\dfrac{\partial (v\rho)}{\partial x}.
\eqno {(10)}
$$

Following step to a method consecutive approximations we search for the decision 
of the non-uniform system of equations (8) -- (10).

The structure of the partial decision of non-uniform system of the equations (8) -- (10) 
has a following appearance, defined by the form (7) of these heterogeneities 
$$
v=v_\infty+v_0e ^ {-2kx} +v_1e ^ {2\xi} +v_1 ^*e ^ {2\xi ^ *},
$$
$$
T=T_\infty + T_0e ^ {-2kx} +T_1e ^ {2\xi} +T_1 ^*e ^ {2\xi ^ *},
$$
$$
\rho =\rho_\infty + \rho_0e ^ {-2kx} + \rho_1e ^ {2\xi} +T_1 ^*e ^ {2\xi ^ *}.
$$

Owing to decrease of speed of environment far from a wall at once it is received, that
$
v_\infty=0.
$
We interest the jump temperature  only. This quantity
does not depend on time. The contribution to it is brought only by quantities
$U_0, v_0, T_0, \rho_0$. These quantities are constant in time. 
Thus $v_0=0$, $ {\rm div} {\bf V} _0=0$, $p =\const $.

Expressions $v\frac{\partial v}{\partial x} $ and $v\frac{\partial s}{\partial x} $ 
have higher order малости on $ \varepsilon $, rather than the second. 
In the same approximation $ \varkappa $ it is possible to consider heat conductivity 
coefficient as constant.

As a result the system of the equations (8) -- (10) is reduced to one non-trivial equation
$$
\varkappa \triangle T +\eta\Big (\dfrac {\partial u} {\partial x} \Big) ^2=0.
\eqno {(11)}
$$

Taking into account a relation (7) from the equation (11) it is received
$$
4 k^2\varkappa T_0 =-k^2\eta U_0^2.
$$
From here we find $T_0$
$$
T_0 = - {\eta U_0^2}/4\varkappa.
$$
Let us designate surface temperature through $T_s $. It is obvious, that
$$
T_s=T_\infty + T_0.
$$
From here follows, that a difference of temperatures between temperature of a surface and
in temperature far from a wall, (i.e. temperature jump $ \delta T $) it is equal
$$
\delta T=T_\infty-T_s =\dfrac {\eta U_0^2} {4\varkappa}.
\eqno {(12)}
$$

Let us notice, that the received expression (12) did not include coefficient
of the second viscosity $ \zeta $. Temperature jump is proportional to coefficient
of dynamic viscosity, square of amplitude of speed of a wall, and it is inverse
proportional to the coefficient of heat conductivity.

The received result is fair both for gas, and for the liquid environment.

Thus, generating of shift waves in the viscous environment by oscillation surface
leads to occurrence of jump of temperature (12) which comes to light as the result
of nonlinear analysis.

Let us notice, that the first researches of jump of temperature in the rarefied gas
were spent in the end of XIX century \cite{Smoluchowski}.

\end{document}